\documentclass{ws-rv961x669}
\usepackage{ws-rv-van}     
\usepackage{color}

\makeindex

\usepackage{hyperref}
\hypersetup{colorlinks=true,citecolor=blue,linkcolor=black,filecolor=black,runcolor=black}

\crop[off]

\begin{document}

\chapter[Pupil Plane Phase]{Pupil Plane Phase Apodization}\label{sec:ppp}

\author[M. Kenworthy, J. Codona and F. Snik]{Matthew A. Kenworthy, Johanan
L. Codona\footnote{Steward Observatory, 933 N. Cherry Avenue, Tucson, AZ
85721, USA} and Frans Snik}

\address{Leiden Observatory, Leiden University, \\
P.O. Box 9513, 2300 RA Leiden, The Netherlands, \\
kenworthy@strw.leidenuniv.nl}

\begin{abstract}

Phase apodization coronagraphs are implemented in a pupil plane to
create a dark hole in the science camera focal plane.
They are successfully created as ``Apodizing Phase Plates'' (APPs) using
classical optical manufacturing, and as ``vector-APPs'' using
liquid-crystal patterning with essentially achromatic performance.
This type of coronagraph currently delivers excellent broadband contrast
($\sim$10$^{-5}$) at small angular separations (few $\lambda/D$) at
ground-based telescopes, owing to their insensitivity to tip/tilt
errors.

\end{abstract}

\body

\section{Introduction}

Pupil-plane apodization techniques (amplitude, phase, or complex) differ
from focal plane mask coronagraphs in that they affect all objects in
the field in an identical fashion.
The main goal of such pupil-plane coronagraphs is to enforce dark holes
in the ensuing point spread function (PSF) in which faint companions can
be directly detected and characterized.
Since the star and companion have the same PSF, the halo should be
suppressed while preserving the starlight in the core as much as
possible, {\it i.e.}~a high Strehl ratio PSF.
In this situation, the ``noise'' is governed by the PSF diffraction halo
plus any diffuse background, while the ``signal'' is contained in the
PSF core.

The phase-only ``Apodizing Phase Plate''\cite{Kenworthy07,
Kenworthy13,Kenworthy10a,Kenworthy10b} (APP) coronagraphs have now been
successfully applied on-sky at ground-based telescopes.
The main benefits of APPs include a high contrast inside the dark hole
($\sim$10$^{-4}$--10$^{-6}$), at a small inner working angle $\sim$$1.5
\lambda/D$,  with complete insensitivity to tip/tilt errors (and
partially resolved stellar disks) that usually limit focal-plane
coronagraphs.
This invariance of the PSF additionally enables beam-switching for
thermal background removal, and observations of multiple star systems.
With the introduction of advanced liquid-crystal technology for the
vector-APP coronagraph\cite{vAPP,vAPP-prototype,vAPP-MagAO}, it has also
become efficient over spectral bandwidths of more than an octave, at
wavelengths from 300 to 30,000 nm\cite{Packham2010}.
The extreme phase patterns enabled by liquid-crystal writing techniques
can now also produce dark holes with various shapes, including
complementary 180$^\circ$ D-shaped dark holes and 360$^\circ$
donut-shaped dark holes.
As a single pupil-plane optic, the (vector-)APP is easily implemented in
a filter wheel in existing instruments, and is fully compatible with
cryovacuum (and likely also space-based) operation.

\section{Theory}

The 1-D apodization problem has been studied for a long time, including
slit apodization in spectroscopy and pulse shaping to reduce channel
bandwidth in telegraphy, by apodizing in amplitude\cite{Jacquinot64}.
The family of functions to describe this are the Slepian functions and
the Prolate Spheroidal wavefunction\cite{slepian1965ast}.
Since transmission apodization is linear, it can achieve a high degree
of suppression between the PSF and the halo beyond a selected inner
working angle (IWA), and in general the apodizations are complex with
both transmission and phase.
The accurate manufacture of complex amplitude masks is non-trivial and
can result in low transmission efficiencies.

Phase-only apodization theory was initially developed for removing
speckles generated by residual optical aberrations in high contrast
imaging experiments\cite{Malbet95}, where wavefront sensing in the final
focal plane of a coronagraph forms a closed loop with a deformable
mirror (DM) in the optical system.
A sinusoidal ripple on the DM forms a diffraction grating in the phase
of the wavefront, generating a pair of speckles that are copies of the
Airy core of the central PSF.
The appropriate choice of spatial phase and amplitude of the ripple
applied to the DM destructively interferes with speckles generated by
aberrations in the optical system.
The same principle can be generalized to cancel out the diffraction
rings of the PSF itself, as demonstrated on-sky by the addition of coma
into an adaptive optics system to cancel out part of the first Airy
ring\cite{Serabyn07}.
Apodization in phase over a two-dimensional region does not yet have an
analytic solution.
Superposing many different phase ripples in the pupil plane to suppress
the diffraction pattern over a region of interest (ROI - typically
defined as a D-shaped region next to the Airy core of the PSF) is
challenging, since the speckles add vectorially  and interfere with each
other, making it a nonlinear problem.
Ref.~\refcite{Codona2007} searched for phase-only apodization solutions
through a modal basis approach.  An ROI is defined in a complex
amplitude focal plane, where the diffraction halo is to be minimized.
A complex amplitude field is defined in the pupil plane, and a Fourier
imaging operator is defined that maps from the pupil plane into the ROI.
Singular Value Decomposition of this operator produces a modal basis set
of complex pupil amplitudes, ordered canonically from the most power
contained within the ROI to the least.
These modes typically have complex amplitudes in the pupil plane, so
their complex amplitude is normalized to unity to make them phase-only
apodization.
These ``antihalo'' modes are subtracted off the complex amplitude of the
pupil plane, and the process is repeated.
The antihalo modes extend a short distance beyond the ROI, and if the
IWA is within the first Airy ring, flux from the core of the PSF is
detrimentally removed as well.
Care is needed to suppress these modes by imposing additional
constraints to maximize the PSF core encircled energy.
If not properly accounted for, phase wrapping can also occur when the
peak-to-valley phase apodization is greater than $2\pi$.

New algorithms have been developed at Leiden Observatory by Doelman,
Keller and Por.
Doelman generates focal plane dark zones using a combination of
phase-only pupil modes.\cite{Doelman16}
A simulated annealing approach is used, where the mode amplitudes are
randomly adjusted.
Solutions that improve the dark region are kept, but worse solutions are
occasionally accepted as well to escape local minima.
Keller uses a Gerchberg-Saxton\cite{Gerchberg72} method, switching
between the pupil plane and focal plane.
Convergence to a given contrast level is increased by an order of
magnitude using Douglas-Rachford operator splitting\cite{Douglas56}.
Por\cite{por2017optimal} generalizes an algorithm by
Carlotti\cite{Carlotti2013} for general complex amplitudes in the pupil
plane.
Strehl ratio maximisation for this mask is a linear operation solved by
large scale optimizer, and phase-only solutions are naturally found
through this approach.

\section{First generation APPs using classical phase}

The manufacture of APP solutions requires the variation of phase across
the pupil plane of the camera, and the development of free-form optic
manufacture with notable departures from sphericity using
computer-controlled diamond turning\cite{Davis07} encoded the phase
patterns as variations in the thickness of a high refractive index
transmissive substrate.
First light observations of an APP with diamond turned
optics\cite{Kenworthy07} demonstrated the viability of the manufacturing
technique and of the theory.
The success of the prototype led to APP coronagraphs on the 6.5m MMTO
telescope in Arizona\cite{Kenworthy13} and on the Very Large Telescope
in Chile\cite{Kenworthy10a,Kenworthy10b}.
The VLT APP led to the first coronagraphic image of the extrasolar
planet $\beta$ Pictoris b\cite{Quanz10} and the discovery of the
extrasolar planet HD~100546b\cite{Quanz12}.

Diamond turning only allows for low spatial frequencies in the azimuthal
direction of the cutting tip, and the classical phase plate
manufacturing was inherently chromatic.
Attempts to achromatize the APP using doublets proved highly
challenging\cite{Kenworthy10c}.

\section{The Vector-APP}

The main limitations of the APP coronagraph (chromaticity, limited
coverage around the star, limited phase pattern accuracy) were solved by
the introduction of the vector-APP (vAPP)\cite{vAPP}.
In a similar way as for the Vector Vortex Coronagraph\cite{VVC}, the
vAPP replaces the classical phase pattern ($\phi_{\textrm{c}}[u,v] =
n(\lambda) \Delta d[u,v]$) with the so-called
Pancharatnam\cite{Pancharatnam}-Berry\cite{Berry} phase or ``geometric
phase''\cite{Escuti-geometricphase}.
The vAPP phase pattern is imposed by a half-wave retarder with a
patterned fast axis orientation $\theta[u,v]$.
The geometric phase is imprinted on incident beams decomposed according
to circular polarization state: $\phi_{\textrm{g}}[u,v] =
\pm2\cdot\theta[u,v]$, with the sign depending on the circular
polarization handedness.
As this fast axis orientation pattern does not vary as a function of
wavelength (with the possible exception of an inconsequential
offset/piston term), the geometric phase is strictly achromatic.
A simple Fraunhofer propagation from the pupil $[u,v]$ to the focal
plane $[x,y]$ shows that after splitting circular polarization states
the two ensuing coronagraphic PSFs are point-symmetric
($PSF_{\textrm{L}}[x,y] = PSF_{\textrm{R}}[-x,-y]$), and therefore, in
the case of D-shaped dark holes, delivers complementary PSFs that
furnish instantaneous 3$60^\circ$ search space around each star.

Vector-APP devices are produced by applying two breakthrough
liquid-crystal techniques: any desired phase pattern is applied onto a
substrate glass through a \textit{direct-write
procedure}\cite{directwrite} that applies the orientation pattern
$\theta[u,v]$ by locally polymerizing the alignment layer material in
the direction set by the controllable polarization of a scanning UV
laser.
Consecutively, birefringent liquid-crystal layers are deposited on top
of this alignment layer.
Several self-aligning layers (``\textit{Multi-Twist Retarders}'';
MTR\cite{MTR}) with predetermined parameters (birefringence dispersion,
thickness, nematic twist) yield a linear retardance that is close to
half-wave over the specified wavelength range.
The vAPP can become efficient over a large wavelength range (up to more
than an octave), while any phase pattern can be written with high
accuracy.

\subsection{Prototyping and first on-sky results}

The first broad-band vAPP device was fully characterized in the lab at
visible wavelengths (500--900 nm)\cite{vAPP-prototype}.
The main limitation of the contrast performance inside the dark hole was
the occurrence of leakage terms that produced a faint copy of the
regular PSF on top of the coronagraphic PSFs.
These leakage terms are caused by small offsets to the half-wave
retardance of the vAPP device, and offsets from quarter-wave retardance
of the quarter-wave plate that, together with a Wollaston prism,
accomplishes the (broad-band) circular polarization splitting.
This issue was resolved with the introduction of the
``grating-vAPP''\cite{grating-vAPP}, which implements the circular
polarization splitting by superimposing a tilt (i.e.~a ``polarization
grating''\cite{Packham2010}) pattern on top of the coronagraphic pupil
phase pattern, which, by virtue of the properties of the geometric
phase, very efficiently sends the coronagraphic PSFs into grating orders
$\pm$1, and leaves all the leakage terms in the zeroth order.
The grating-vAPP also greatly simplifies the optical configuration, as
all the manipulation takes place within one single (flat) optic.
The coronagraphic PSFs are now subject to a lateral grating dispersion
term and so the grating-vAPP can only be used in combination with
narrow-band filters, although the wavelength range throughout which
these filters can be applied can still be very large.

\begin{figure}[ht] \includegraphics[width=\textwidth]{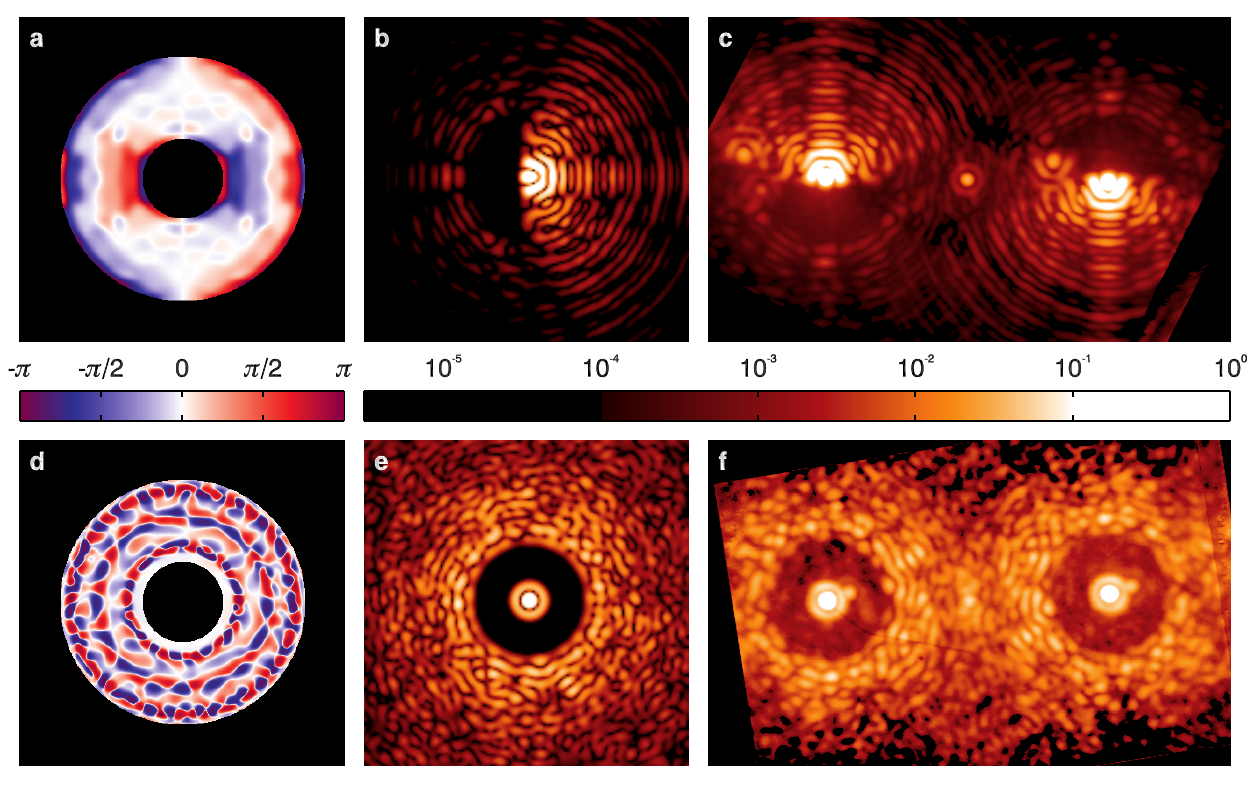}

 \caption{Phase patterns, theoretical and on-sky PSFs (logarithmic
scale) for the two vAPP devices installed at MagAO.
a) Theoretical phase pattern for a 180$^\circ$ dark hole covering 2--7
$\lambda/D$, b) the ensuing theoretical PSF, c) the on-sky PSFs at MagAO
for the star $\eta$ Crucis at 3.9 $\mu$m.
d) Theoretical phase pattern for a 360$^\circ$ dark hole covering 3--7
$\lambda/D$, e) the ensuing theoretical PSF, f) the on-sky PSFs at MagAO
for the binary star $\beta$ Centauri at 3.9 $\mu$m.
Phase pattern designs by Christoph Keller. Data processing by Gilles
Otten\cite{vAPP-MagAO}.}

\label{MagAO-vAPPs}
\end{figure}

The first grating-vAPP successfully demonstrated on-sky was installed at
the MagAO/Clio instrument attached to the 6.5-m Magallan-Clay telescope
in Chile\cite{vAPP-MagAO} (\fref{MagAO-vAPPs}a--c).
The device was designed and built to operate from 2--5 $\mu$m, covering
the infrared atmospheric K, L and M-bands.
The first-light observations demonstrated excellent suppression of the
stellar diffraction halo in the complementary dark holes (see
\fref{MagAO-vAPPs}c).
Detailed analysis of the data demonstrated a 5-$\sigma$ contrast for
point source detection of $\sim$$10^{-5}$ at 2.5--7
$\lambda/D$\cite{vAPP-MagAO}.
The contrast performance is greatly enhanced by combining the two
complementary dark holes through a simple rotation-subtraction procedure
to further suppress the wind-driven starlight halo in the dark holes,
which is caused by finite AO loop speed.
\Fref{MagAO-vAPPs}c shows the presence of the leakage term PSF in
between the coronagraphic PSFs, which can be used as an astrometric and
photometric reference, in the (frequent) case that the coronagraphic PSF
cores are saturated.

\subsection{360 degree APP solutions}

As part of the algorithm exploration of the APP surface, a family of
functions was found that showed 360 degrees of suppression around the
central star.
These solutions have lower Strehl ratios for the star (typically
20--40\%) with larger IWA compared to the 180$^\circ$ dark holes, and
these phase pattern solutions are pathological in nature, with rapid
phase changes over small scales.
The advent of liquid-crystal patterning encouraged us to revisit these
360$^\circ$ solutions, and test them in the lab and on-sky.
\Fref{MagAO-vAPPs}d-f shows the phase pattern and ensuing PSFs for the
experimental vAPP device at MagAO.
The lower row of figures shows that the liquid-crystal manufacture
successfully reproduces the complex phase pattern, and this on-sky image
(\fref{MagAO-vAPPs}f) shows a fainter binary stellar companion to the
right of the primary star's PSF.

\section{Future Directions}

Our team is currently installing different vAPP coronagraphs at several
instruments at large telescopes around the world, and working on novel
designs for the future extremely large telescopes.
Foreseeable future developments of the vector-APP as a separate optical
component, and as integral part of a high-contrast imaging system
include:

\begin{itemize}

\item The combination of several grating layers in a
``\textit{double-grating-vAPP}'' to recombine the two coronagraphic PSFs
with 360$^\circ$ dark holes to feed an integral-field unit while
rejecting the leakage terms. 

\item By prescribing a specific retardance profile as a function of
wavelength, it is possible to build a \textit{wavelength-selective vAPP}
device, that operates as a regular vAPP coronagraph at the science
wavelengths, and acts like a regular glass plate at the spectral range
of a wavefront sensor behind it. 

\item The pupil phase manipulation of the vAPP can be extended by
amplitude manipulation in the pupil to create complex
apodizers\cite{Carlotti2013}, and by phase/amplitude masks in the focal
plane to yield \textit{hybrid coronagraphy}\cite{Ruane2015}.

\item As this technology is likely compatible with operation in space,
it is opportune to characterize the performance of vAPP-like
coronagraphs at the \textit{extreme contrast levels} ($\sim$10$^{-9}$)
of space-based high-contrast imaging.

\item To adapt the vAPP phase pattern to the observational needs, the
observing conditions, and segmented pupils with variable configurations,
active liquid-crystal devices will be developed to establish
``\textit{adaptive coronagraphy}''. Such a system can then deliver dark
holes of various geometry and depth, depending on whether the observer
is interested in detecting exoplanets or characterizing known targets.

\item As the vAPP relies on polarization splitting, it is possible to
design an optimal system for \textit{coronagraphic
polarimetry}\cite{vAPP-polarimetry}, particularly with the
360$^\circ$-designs.

\item The fact that the vAPP produces several PSFs for the same star at
the focal plane makes it an attractive option for implementing
\textit{focal-plane wavefront sensing}, for instance through
phase-diversity techniques. Another promising approach involves the
incorporation of an additional pupil phase pattern which generates pairs
of PSF copies around the main PSFs, with each pair encoding a wavefront
error mode through an intensity difference\cite{Wilby2016}.

\end{itemize}

\section{Acknowledgments}

The research of FS leading to these results has received funding from
the European Research Council under ERC Starting Grant agreement 678194
(FALCONER).

\bibliographystyle{ws-rv-van}
\bibliography{kenworthy_handbook}

\begin{thebibliography}{32}
\providecommand{\natexlab}[1]{#1}
\providecommand{\url}[1]{\texttt{#1}}
\expandafter\ifx\csname urlstyle\endcsname\relax
  \providecommand{\doi}[1]{doi: #1}\else
  \providecommand{\doi}{doi: \begingroup \urlstyle{rm}\Url}\fi

\bibitem{Kenworthy07}
M.~A. {Kenworthy}, J.~L. {Codona}, P.~M. {Hinz}, J.~R.~P. {Angel}, A.~{Heinze},
  and S.~{Sivanandam}, {First On-Sky High-Contrast Imaging with an Apodizing
  Phase Plate}, \emph{\apj}. {\bf 660}, \penalty0 762--769 (May, 2007).
\newblock \doi{10.1086/513596}.

\bibitem{Kenworthy13}
M.~A. {Kenworthy}, T.~{Meshkat}, J.~H. {Quanz}, S.~P. and·~{Girard}, M.~R.
  {Meyer}, and M.~{Kasper}, {Coronagraphic Observations of Fomalhaut at Solar
  System Scales}, \emph{\apj}. 764:\penalty0 7 (Feb., 2013).
\newblock \doi{10.1088/0004-637X/764/1/7}.

\bibitem{Kenworthy10a}
M.~{Kenworthy}, S.~{Quanz}, M.~{Meyer}, M.~{Kasper}, J.~{Girard}, R.~{Lenzen},
  J.~{Codona}, and P.~{Hinz}, {A New Coronagraph for NAOS-CONICA -- the
  Apodising Phase Plate}, \emph{The Messenger}. {\bf 141}, \penalty0 2--4 (sep,
  2010).

\bibitem{Kenworthy10b}
M.~A. {Kenworthy}, S.~P. {Quanz}, M.~R. {Meyer}, M.~E. {Kasper}, R.~{Lenzen},
  J.~L. {Codona}, J.~H. {Girard}, and P.~M. {Hinz}.
\newblock {An apodizing phase plate coronagraph for VLT/NACO}.
\newblock In \emph{Society of Photo-Optical Instrumentation Engineers (SPIE)
  Conference Series}, vol. 7735, \emph{Presented at the Society of
  Photo-Optical Instrumentation Engineers (SPIE) Conference} (July, 2010).
\newblock \doi{10.1117/12.856811}.

\bibitem{vAPP}
F.~{Snik}, G.~{Otten}, M.~{Kenworthy}, M.~{Miskiewicz}, M.~{Escuti},
  C.~{Packham}, and J.~{Codona}.
\newblock {The vector-APP: a broadband apodizing phase plate that yields
  complementary PSFs}.
\newblock In \emph{Proc.~SPIE}, vol. 8450, \emph{Proc.~SPIE} (Sept., 2012).
\newblock \doi{10.1117/12.926222}.

\bibitem{vAPP-prototype}
G.~P. P.~L. Otten, F.~Snik, M.~A. Kenworthy, M.~N. Miskiewicz, and M.~J.
  Escuti, Performance characterization of a broadband vector apodizing phase
  plate coronagraph, \emph{Opt. Express}. {\bf 22}\penalty0 (24), \penalty0
  30287--30314 (Dec, 2014).
\newblock \doi{10.1364/OE.22.030287}.
\newblock URL \url{http://www.opticsexpress.org/
  abstract.cfm?URI=oe-22-24-30287}.

\bibitem{vAPP-MagAO}
G.~P.~P.~L. {Otten}, F.~{Snik}, M.~A. {Kenworthy}, C.~U. {Keller}, J.~R.
  {Males}, K.~M. {Morzinski}, L.~M. {Close}, J.~L. {Codona}, P.~M. {Hinz},
  K.~J. {Hornburg}, L.~L. {Brickson}, and M.~J. {Escuti}, {On-sky Performance
  Analysis of the Vector Apodizing Phase Plate Coronagraph on MagAO/Clio2},
  \emph{\apj}. 834:\penalty0 175 (Jan., 2017).
\newblock \doi{10.3847/1538-4357/834/2/175}.

\bibitem{Packham2010}
C.~{Packham}, M.~{Escuti}, J.~{Ginn}, C.~{Oh}, I.~{Quijano}, and G.~{Boreman},
  {Polarization Gratings: A Novel Polarimetric Component for Astronomical
  Instruments}, \emph{PASP}. {\bf 122}, \penalty0 1471--1482 (Dec., 2010).
\newblock \doi{10.1086/657904}.

\bibitem{Jacquinot64}
P.~{Jacquinot} and B.~{Roizen-Dossier}, \emph{{Progress in Optics}}, vol.~3,
  chapter~2, pp. 29--186.
\newblock North Holland Publishing Company,  (1964).

\bibitem{slepian1965ast}
D.~Slepian, {Analytic solution of two apodization problems}, \emph{J. Opt. Soc.
  Am}. {\bf 55}\penalty0 (9), \penalty0 1110--1115,  (1965).

\bibitem{Malbet95}
F.~{Malbet}, J.~W. {Yu}, and M.~{Shao}, {High-Dynamic-Range Imaging Using a
  Deformable Mirror for Space Coronography}, \emph{\pasp}. {\bf 107}, \penalty0
  386 (Apr., 1995).
\newblock \doi{10.1086/133563}.

\bibitem{Serabyn07}
E.~{Serabyn}, K.~{Wallace}, M.~{Troy}, B.~{Mennesson}, P.~{Haguenauer},
  R.~{Gappinger}, and R.~{Burruss}, {Extreme Adaptive Optics Imaging with a
  Clear and Well-Corrected Off-Axis Telescope Subaperture}, \emph{\apj}. {\bf
  658}, \penalty0 1386--1391 (Apr., 2007).
\newblock \doi{10.1086/511949}.

\bibitem{Codona2007}
J.~{Codona}.
\newblock {Phase Apodization Coronagraphy}.
\newblock In \emph{In the Spirit of Bernard Lyot: The Direct Detection of
  Planets and Circumstellar Disks in the 21st Century}, p.~24 (June, 2007).

\bibitem{Doelman16}
D.~{Doelman}.
\newblock {Optimizing apodizing phase plate designs with simulated annealing}.
\newblock Master's thesis, Leiden Observatory, {Leiden, The Netherlands},
  (2016).

\bibitem{Gerchberg72}
R.~W. Gerchberg and W.~O. Saxton, {A practical algorithm for the determination
  of the phase from image and diffraction plane pictures}, \emph{Optik (Jena)}.
  {\bf 35}, \penalty0 237+,  (1972).

\bibitem{Douglas56}
J.~Douglas and H.~Rachford, On the numerical solution of heat conduction
  problems in two or three space variables, \emph{Trans. Amer. Math. Soc.} {\bf
  82}\penalty0 (2), \penalty0 421--439,  (1956).

\bibitem{por2017optimal}
E.~H. {Por}.
\newblock {Optimal design of apodizing phase plate coronagraphs}.
\newblock In \emph{Society of Photo-Optical Instrumentation Engineers (SPIE)
  Conference Series}, vol. 10400, p. 104000V (Sept., 2017).
\newblock \doi{10.1117/12.2274219}.

\bibitem{Carlotti2013}
A.~{Carlotti}, N.~J. {Kasdin}, R.~J. {Vanderbei}, and A.~J. {Eldorado Riggs}.
\newblock {Hybrid coronagraphic design: optimization of complex apodizers}.
\newblock In \emph{Proc.~SPIE}, vol. 8864, \emph{Proc.~SPIE} (Sept., 2013).
\newblock \doi{10.1117/12.2024523}.

\bibitem{Davis07}
G.~E. {Davis}, M.~A. {Kenworthy}, and A.~R. {Hedges}.
\newblock {Manufacturing of a freeform phase plate for suppression of
  diffraction in an astronomical telescope}.
\newblock In \emph{Proc. SPIE TD04, Optifab 2007: Technical Digest, TD041J;
  James J. Kumler; Matthias Pfaff, Eds.} (May, 2007).

\bibitem{Quanz10}
S.~P. {Quanz}, M.~R. {Meyer}, M.~A. {Kenworthy}, J.~H.~V. {Girard},
  M.~{Kasper}, A.-M. {Lagrange}, D.~{Apai}, A.~{Boccaletti}, M.~{Bonnefoy},
  G.~{Chauvin}, P.~M. {Hinz}, and R.~{Lenzen}, {First Results from Very Large
  Telescope NACO Apodizing Phase Plate: 4 {$\mu$}m Images of The Exoplanet
  {$\beta$} Pictoris b}, \emph{\apjl}. {\bf 722}, \penalty0 L49--L53 (Oct.,
  2010).
\newblock \doi{10.1088/2041-8205/722/1/L49}.

\bibitem{Quanz12}
S.~P. {Quanz}, J.~R. {Crepp}, M.~{Janson}, H.~{Avenhaus}, M.~R. {Meyer}, and
  L.~A. {Hillenbrand}, {Searching for Young Jupiter Analogs around AP Col:
  L-band High-contrast Imaging of the Closest Pre-main-sequence Star},
  \emph{\apj}. 754:\penalty0 127 (Aug., 2012).
\newblock \doi{10.1088/0004-637X/754/2/127}.

\bibitem{Kenworthy10c}
M.~A. {Kenworthy}, P.~M. {Hinz}, J.~L. {Codona}, J.~C. {Wilson}, M.~F.
  {Skrutskie}, and E.~{Solheid}.
\newblock {Developing achromatic coronagraphic optics for LMIRCam and the LBT}.
\newblock In \emph{Society of Photo-Optical Instrumentation Engineers (SPIE)
  Conference Series}, vol. 7734, \emph{Presented at the Society of
  Photo-Optical Instrumentation Engineers (SPIE) Conference} (July, 2010).
\newblock \doi{10.1117/12.856819}.

\bibitem{VVC}
D.~{Mawet}, E.~{Serabyn}, K.~{Liewer}, C.~{Hanot}, S.~{McEldowney}, D.~{Shemo},
  and N.~{O'Brien}, {Optical Vectorial Vortex Coronagraphs using Liquid Crystal
  Polymers: theory, manufacturing and laboratory demonstration}, \emph{Optics
  Express}. {\bf 17}, \penalty0 1902--1918 (Feb., 2009).
\newblock \doi{10.1364/OE.17.001902}.

\bibitem{Pancharatnam}
S.~Pancharatnam, Generalized theory of interference, and its applications. part
  i. coherent pencils, \emph{Proceedings of the Indian Academy of Sciences,
  Section A}. {\bf 44}\penalty0 (5), \penalty0 247--262,  (1956).

\bibitem{Berry}
M.~V. Berry, Quantal phase factors accompanying adiabatic changes,
  \emph{Proceedings of the Royal Society of London A: Mathematical, Physical
  and Engineering Sciences}. {\bf 392}\penalty0 (1802), \penalty0 45--57,
  (1984).
\newblock ISSN 0080-4630.
\newblock \doi{10.1098/rspa.1984.0023}.
\newblock URL \url{http://rspa.royalsocietypublishing.org/content/392/1802/45}.

\bibitem{Escuti-geometricphase}
M.~J. Escuti, J.~Kim, and M.~W. Kudenov, Controlling light with geometric-phase
  holograms, \emph{Opt. Photon. News}. {\bf 27}\penalty0 (2), \penalty0 22--29
  (Feb, 2016).
\newblock \doi{10.1364/OPN.27.2.000022}.
\newblock URL \url{http://www.osa-opn.org/abstract.cfm?URI=opn-27-2-22}.

\bibitem{directwrite}
M.~N. Miskiewicz and M.~J. Escuti, Direct-writing of complex liquid crystal
  patterns, \emph{Opt. Express}. {\bf 22}\penalty0 (10), \penalty0 12691--12706
  (May, 2014).
\newblock \doi{10.1364/OE.22.012691}.
\newblock URL \url{http://www.opticsexpress.org/
  abstract.cfm?URI=oe-22-10-12691}.

\bibitem{MTR}
R.~K. Komanduri, K.~F. Lawler, and M.~J. Escuti, Multi-twist retarders:
  broadband retardation control using self-aligning reactive liquid crystal
  layers, \emph{Opt. Express}. {\bf 21}\penalty0 (1), \penalty0 404--420 (Jan,
  2013).
\newblock \doi{10.1364/OE.21.000404}.
\newblock URL \url{http://www.opticsexpress.org/abstract.cfm?URI=oe-21-1-404}.

\bibitem{grating-vAPP}
G.~P. P.~L. Otten, F.~Snik, M.~A. Kenworthy, M.~N. Miskiewicz, M.~J. Escuti,
  and J.~L. Codona.
\newblock The vector apodizing phase plate coronagraph: prototyping,
  characterization and outlook.
\newblock In \emph{Society of Photo-Optical Instrumentation Engineers (SPIE)
  Conference Series}, vol. 9151, pp. 91511R--91511R--10,  (2014).
\newblock \doi{10.1117/12.2056096}.
\newblock URL \url{http://dx.doi.org/10.1117/12.2056096}.

\bibitem{Ruane2015}
G.~J. {Ruane}, E.~{Huby}, O.~{Absil}, D.~{Mawet}, C.~{Delacroix},
  B.~{Carlomagno}, and G.~A. {Swartzlander}, {Lyot-plane phase masks for
  improved high-contrast imaging with a vortex coronagraph}, \emph{A\&A}.
  583:\penalty0 A81 (Nov., 2015).
\newblock \doi{10.1051/0004-6361/201526561}.

\bibitem{vAPP-polarimetry}
F.~{Snik}, G.~{Otten}, M.~{Kenworthy}, D.~{Mawet}, and M.~{Escuti}.
\newblock {Combining vector-phase coronagraphy with dual-beam polarimetry}.
\newblock In \emph{Ground-based and Airborne Instrumentation for Astronomy V},
  vol. 9147, \emph{\procspie}, p. 91477U (Aug., 2014).
\newblock \doi{10.1117/12.2055452}.

\bibitem{Wilby2016}
M.~J. {Wilby}, C.~U. {Keller}, F.~{Snik}, V.~{Korkiakoski}, and A.~G.~M.
  {Pietrow}, {The coronagraphic Modal Wavefront Sensor: a hybrid focal-plane
  sensor for the high-contrast imaging of circumstellar environments},
  \emph{\aap}. 597:\penalty0 A112 (Jan., 2017).
\newblock \doi{10.1051/0004-6361/201628628}.

\end{thebibliography}
\end{document}